# Identifying the origin of the non-monotonic thickness dependence of spin-orbit torques and interfacial Dzyaloshinskii-Moriya interaction in a ferrimagnetic insulator heterostructure

*Shilei Ding* [1,2,3], *Lorenzo Baldrati* [2], *Andrew Ross* [2,3], *Zengyao Ren* [2,3,4], *Rui Wu* [2,7], *Sven Becker* [2], *Jinbo Yang* [1,5,6], *Gerhard Jakob* [2,3], *Arne Brataas*[7], *Mathias Kläui* [2,3,7*]

[1]State Key Laboratory for Mesoscopic Physics, School of Physics, Peking University, Beijing 100871, China.

[2]Institute of Physics, Johannes Gutenberg-University Mainz, Staudingerweg 7, 55128 Mainz, Germany.

[3]Graduate School of Excellence Materials Science in Mainz, 55128 Mainz, Germany.

[4]School of Materials Science and Engineering, University of Science and Technology Beijing, Beijing 100083, China.

[5]Collaborative Innovation Center of Quantum Matter, Beijing, 100871, P.R. China.

[6]Beijing Key Laboratory for Magnetoelectric Materials and Devices, Beijing 100871, P. R. China.

[7]Center for Quantum Spintronics, Department of Physics, Norwegian University of Science and Technology, NO-7491 Trondheim, Norway.

*Correspondence to [klaeui@uni-mainz.de]

**Abstract**

Electrical manipulation of magnetism via spin-orbit torques (SOTs) promises efficient spintronic devices. In systems comprising magnetic insulators and heavy metals, SOTs have started to be investigated only recently, especially in systems with interfacial Dzyaloshinskii-Moriya interaction (iDMI). Here, we quantitatively study the SOT efficiency and iDMI in a series of gadolinium gallium garnet (GGG) / thulium iron garnet (TmIG) / platinum (Pt) heterostructures with varying TmIG and Pt thicknesses. We find that the non-monotonic SOT efficiency as a function of the magnetic layer thickness is not consistent with the 1/thickness dependence expected from a simple interfacial SOT mechanism. Moreover, considering the insulating nature of TmIG, our results cannot be explained by the SOT mechanism established for metallic magnets where the transverse charge spin current can inject and dephase in the magnetic layers. Rather we can explain this non-monotonic behavior by a model based on the interfacial spin mixing conductance that is affected by the thickness-dependent exchange splitting energy by determining the phase difference of the reflected spin-up and spin-down electrons at the TmIG / Pt interface. By studying the Pt thickness dependence, we find that the effective DMI for GGG / TmIG / Pt does not depend on the Pt thickness, which indicates that the GGG / TmIG interface is the source of the iDMI in this system. Our work demonstrates that SOT and DMI can originate from two different interfaces, which enables independent optimization of DMI and SOT for advanced chiral spintronics with low damping magnetic insulators.

**Keywords:** magnetic insulator; spin-orbit torque; Dzyaloshinskii-Moriya interaction; spin mixing conductance

Electrical manipulation of the magnetization in ferromagnet (FM) / heavy metal (HM) bilayers via spin-orbit torques (SOTs) has attracted great interest in the field of spintronics for its potential use in logic and memory applications [1-7]. SOTs [3] occur by a non-equilibrium spin current or spin density generated by the spin Hall effect (SHE) [1,4,5] or the inverse spin galvanic effect (ISGE) [6,7], and these spins then interact with the magnetization. So SOT efficiency depends on the exchange interaction between the spins of the spin current and local magnetic moments. The torque exerted on the magnetic moments can be decomposed into two components with different symmetries, namely the damping-like torque $\tau_{DL}$ proportional to $m \times (m \times \sigma)$ and the field-like torque $\tau_{FL}$ proportional to $m \times \sigma$, where $\sigma$ is the polarization direction of the spin current and $m$ is the magnetization vector in the FM [3]. A spin current of polarization $\sigma$ that is not parallel to $m$ (transverse spin current) can transfer the angular momentum to the magnetization and thereby exert torques. The role of the transverse spin currents in systems comprising ferromagnetic metals (FMM) has been widely studied, for its importance in the spin-transfer torque switching of the magnetization of the free layer in metallic spin-valve devices [8,9], where the spins are carried by electrons. The injected spins interact with the magnetic moments and dephase in the FMM leading to the transfer of angular momentum mentioned above.

The situation is different in ferrimagnetic insulators (FMI), where a spin-polarized charge current cannot flow. In FMI / HM systems, one can drive a charge current in the HM layer and generate a spin accumulation at the FMI / HM interface by the spin Hall effect [1]. A collinear spin current with polarization parallel (antiparallel) to the magnetization can create (annihilate) magnons in the FMI [10]. In the case of a transverse spin current, the electrons in the HM are unable to enter the insulator, however theoretically it is predicted that the exchange splitting energy and spin mixing conductance can influence the magnitude of the SOT generated at the FMI / HM interface [11]. To date, the behavior of a transverse spin current at the FMI / HM interface has not been fully understood, but it is of key importance, as it governs the SOT induced switching of any FMI. The second main ingredient for the magnetic switching, via domain wall motion induced by SOT, is the Dzyaloshinskii-Moriya interaction (DMI), which also results from spin-orbit coupling effects [12,13]. Often, a sizeable DMI is found in systems where SOT are observed, and the DMI stabilizes chiral spin structures such as skyrmions and chiral domain walls [14], which can then be efficiently manipulated by SOT [15,16]. Some models predict that SOT and DMI can be related [17], while experimentally in some metallic systems they are found not to share the same origin [18]. However, in multilayers with ferromagnetic insulators the situation is fundamentally different from metals: While in metals, a charge current can flow across both interfaces of a magnetic layer, in a system with an insulator and a conducting layer, the current can flow only at one interface. This allows one to confine the SOT generation at one interface, while the DMI can be generated at the other interface, and thus potentially both can be tuned independently.

In particular, the magnetic insulator thulium iron garnet ($Tm_3Fe_5O_{12}$;TmIG) was shown to host chiral spin structures that can be manipulated by SOT [19-26]. Moreover, spin Hall magnetoresistance [19,21], efficient magnetization switching [19-21] and current-induced domain wall motion [21-26] have been reported in systems comprising this material. TmIG shows perpendicular magnetization anisotropy (PMA) in a large range of thicknesses from 1.7 nm to at least 15 nm, and, in PMA GGG / TmIG / Pt heterostructures, the robust anomalous Hall effect (AHE) makes it a good candidate to explore the role of SOT at the FMI / HM interface. In addition, it was reported that the DMI in TmIG is of interfacial origin (iDMI) [21, 23-26], with some reporting that the iDMI originates from the TmIG / HM (e.g Pt) interface [23,24], while others claim that the interface between TmIG and the substrate is responsible for the iDMI [21,25,26], so that the origin of the iDMI is currently being debated. It is thus important to understand the origin of the SOT and the DMI in a substrate / FMI / HM structure, where the SOT and DMI could originate from two different interfaces, as this potentially enables also independent optimization of the DMI and SOT in magnetic insulators.

In this work, we perform a systematic thickness dependence study of the SOT efficiency and DMI in GGG / TmIG / HM heterostructures. We find that the SOT efficiency exhibits a non-monotonic dependence on the TmIG thickness ($t_{TmIG}$), with a peak at 5.4 nm. We explain this finding using a model that combines the thickness-dependent magnetic properties of the TmIG in the low thickness region, which influence the spin mixing conductance via the spin splitting of the magnon modes, and the decay expected from an interfacial mechanism for thicker TmIG. Comparing SOT and DMI, we find that the DMI energy density is almost constant as a function of the Pt thickness ($t_{Pt}$), which strongly suggests that the iDMI in this system is not originating from the TmIG / HM interface but is rather determined by the GGG / TmIG interface. Our results show that it is possible to independently optimize the interfacial SOT and DMI in TmIG (see schematic in Figure 1a), which is important in view of device applications.

To obtain high-quality crystalline epitaxial samples, we have grown TmIG films with PMA on (111)-oriented gadolinium gallium garnet ($Gd_3Ga_5O_{12}$, GGG) substrates by pulsed laser deposition.

The deposition conditions have been reported elsewhere [21], The Pt, W, and Cu layers have been deposited subsequently by sputtering in a chamber with a base vacuum pressure of $3 \times 10^{-9}$ mbar, the sputtering Ar pressure is 0.013 mbar. The Pt thickness $t_{Pt}$ is varied in fine steps from 0.5 to 7 nm. In order to study the SOT and DMI, the thin film samples have been patterned and etched into 6 μm × 30 μm Hall bars by photolithography and ion milling. All measurements shown below are carried out at room temperature (RT), and the effective thickness of TmIG is modified by subtracting the magnetic dead layer (details, see supplemental material S1 [27]).

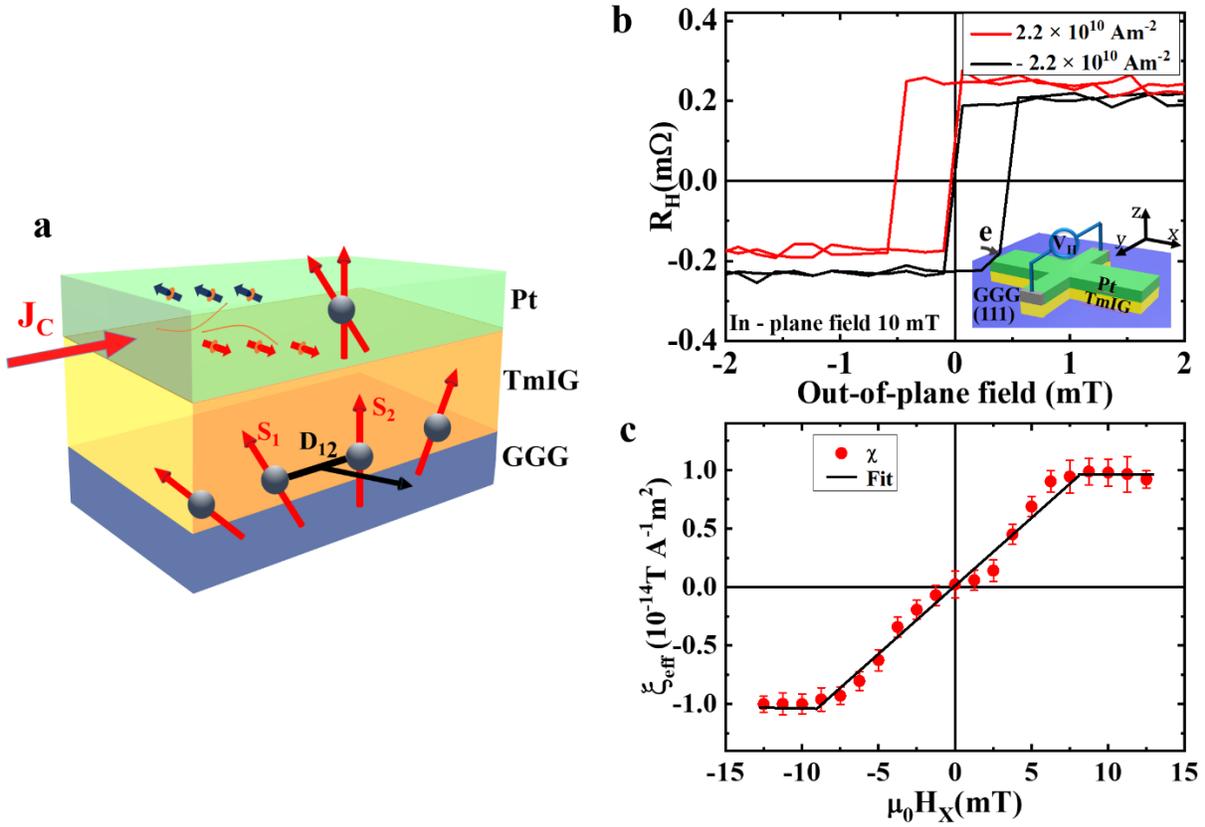

**Figure 1**. (a) The SOT and iDMI can originate from different interfaces in a GGG / TmIG / Pt heterostructure. (b) Horizontal shift of the out-of-plane hysteresis curves measured with an applied in-plane field of $\mu_0 H_x$ =10 mT, slight vertical offsets are introduced for clarity. The probe current density

is $2.2 \times 10^{10}$ Am$^{-2}$ with different polarities. The inset shows the Hall bar schematic. (c) The $\mu_0 H_x$ dependence of the effective SOT efficiency $\xi_{eff}$ for GGG / TmIG (5.4) / Pt (7) (units in nanometer).

First, we start by measuring the hysteresis loop of the Hall resistance $R_H$ as a function of the out-of-plane magnetic field, in the presence of different applied in-plane field $\mu_0 H_x$ and charge current density $j$. In a system with sizable DMI, Néel type domain walls are energetically favorable over the Bloch type domain walls [28]. The current-induced effective field $\mu_0 H_{eff}$, generated by the damping-like spin orbit torque, can drive the Néel domain wall similarly to the effect of an out-of-plane field in a direction related to the chirality of the domain wall [28], which will lead the horizontal shift of the hysteresis loops, similarly to the exchange bias field in PMA exchange bias systems [29]. Figure 1b shows the shifted hysteresis loops in the presence of a charge current of density $+2.2 \times 10^{10}$ Am$^{-2}$ and $-2.2 \times 10^{10}$ Am$^{-2}$ for a GGG / TmIG (5.4) / Pt (7) (units in nanometer) sample, where reversing the polarity of the probing current leads to a negative displacement of the out-of-plane hysteresis. We obtain a current-induced effective damping-like field (measured by the shift of the hysteresis) $\mu_0 H_{eff} = 0.25$ mT for the current density used. The SOT efficiency is defined as at the effective damping-like field, generated via the spin-orbit torque, per unit applied current density $j$, $\xi_{eff} = \partial(\mu_0 H_{eff})/\partial j_c$, as described in Ref. [28]. The $\mu_0 H_x$ dependence of the SOT efficiency is shown in Figure 1c. The slope $\partial(\mu_0 H_{eff})/\partial j_c$ is linear for small $\mu_0 H_x$ before saturating at larger fields, from which we can estimate the saturation SOT efficiency $\xi_{eff} = 1.0 \pm 0.2$ T / $10^{14}$ A m$^{-2}$. We then use the dependence of the SOT efficiency on the in-plane magnetic field to estimate the value of the DMI in this system [28]. The effective DMI field can be estimated from the saturation in-plane field $\mu_0 H_{x-saturation} = \mu_0 H_{DMI} = 9 \pm 1$ mT for GGG / TmIG (5.4) / Pt (7). We compute the effective DMI energy density $|D| = M_s \Delta \mu_0 |H_{DMI}| = 22.7 \pm 4.0$ µJ/m$^2$ (details, see supplemental material S2 [30]).

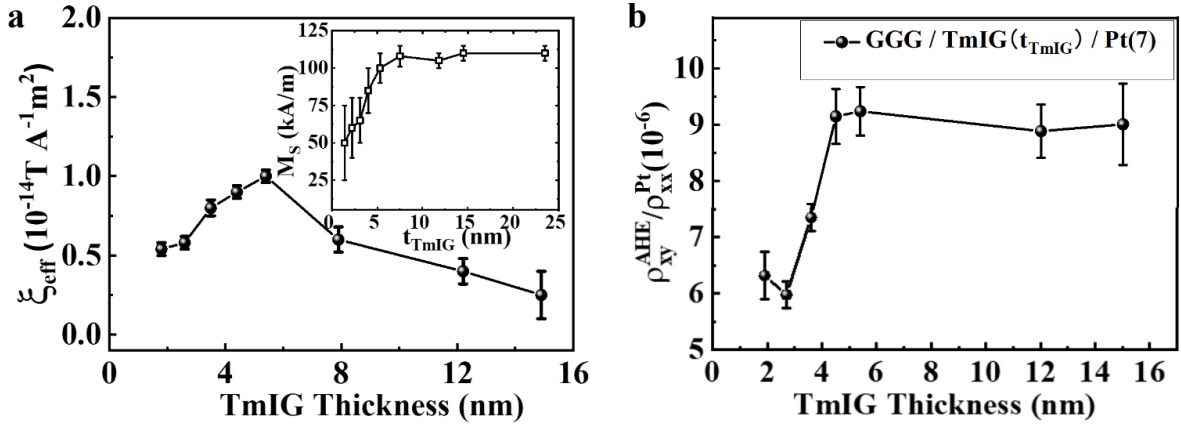

**Figure 2.** (a) SOT efficiency for GGG / TmIG($t_{TmIG}$) / Pt (7) as a function of TmIG thickness $t_{TmIG}$. The inset shows the thickness dependence of saturation magnetization for TmIG. (b) The ratio $\rho_{xy}^{AHE}/\rho_{xx}^{Pt}$ increases with $t_{TmIG}$ which indicates the increase of the spin mixing conductance.

We next consider the behavior of the SOT in a magnetic insulator, for which we perform a thickness dependent study of the SOT efficiency. For most of the studies based on metals, the SOT efficiency $\xi_{eff}$ scales inversely proportional to the thickness of the magnetic material [8,31], as expected from an interface effect. However, we find here a non-monotonic trend for GGG / TmIG ($t_{TmIG}$) / Pt (7) heterostructures with different TmIG thicknesses: $\xi_{eff}$ increases up to around 5.4 nm and decreases with increasing TmIG thickness as shown in Figure 2a. To understand this behavior, we measured the evolution of $M_s$ and the relative amplitude of the AHE resistivity $\rho_{xy}^{AHE}/\rho_{xx}^{Pt}$ as a function of the TmIG thickness in Figure 2b. To extract $\rho_{xy}^{AHE}/\rho_{xx}^{Pt}$, we first quantify the amplitude $R_{xy}^{AHE}$ from the anomalous Hall hysteresis loop, from which we obtain the thickness dependence of the anomalous Hall resistivity $\rho_{xy}^{AHE} = R_{xy}^{AHE} t_{Pt}$. By using the value $\rho_{xx}^{Pt}$ measured on the Hall device, we calculate the ratio $\rho_{xy}^{AHE}/\rho_{xx}^{Pt}$ as a function of the TmIG thickness, as shown in Figure 2b. The amplitude of the AHE in FMI / HM bilayers can depend on two mechanisms, one is the magnetic proximity effect (MPE) in the Pt and the other one is the AHE induced by the spin current *reflection / absorption* related to the spin accumulation induced by the SHE [32]. It has been reported that at a FMI / Pt interface, the magnetic moment induced in the Pt from the magnetic insulator is less than $0.1\mu_B$ [33] and the spin-Hall induced AHE dominates over the AHE induced by the MPE at room temperature in YIG / Pt bilayer system [34]. Given the similarity between YIG and TmIG, it is reasonable to assume that in TmIG / Pt the SHE-induced AHE also dominates at room temperature, implying that the magnitude of $\rho_{xy}^{AHE}/\rho_{xx}^{Pt}$ is a good indicator for the magnitude of the imaginary part of the spin mixing conductance [35]. According to this interpretation, the $t_{TmIG}$ dependence of $\rho_{xy}^{AHE}/\rho_{xx}^{Pt}$ indicates that the spin mixing conductance increases with $t_{TmIG}$ for low TmIG thicknesses. Also note that in the Supplemental material Figure S3 we show that the effective spin Hall angle increases for low TmIG thickness and then plateaus, which should not be interpreted as resulting from a change in the intrinsic spin Hall angle of Pt, but rather as an increasing spin mixing conductance for increasing thickness.

With these experimental observations, we then develop a model to explain our experimental results. It has been reported that the increase of $\xi_{eff}$ is influenced by the $M_s$ of TmIG in the thinner TmIG region [20], where the authors of Ref. [20] claim that $\xi_{eff}$ is proportional to $M_s$. We find that our thickness dependence of $M_s$ of TmIG shown in the inset of Figure 2a is consistent with this previous report. In Ref. [20] the dependence of $\xi_{eff}$ on $M_s$ was obtained phenomenologically, by assuming the expression of the s-d exchange interaction. Here, we provide a model that can as well explain the results, where we focus on how the spin mixing conductance can determine the spin torque efficiency, noting that it is indirectly linked to the magnetization via the exchange splitting of the magnon modes. In fact, the spin mixing conductance depends on the thickness $t_{TmIG}$ through the phase shift of the reflected electrons acquired inside the magnetic insulator. This dependence is controlled by the exchange potentials for spin-up and spin-down electrons [11]. In a simple model, the dimensionless spin mixing conductance for a single conducting channel is $G^{\uparrow\downarrow} = 1 - r^{\uparrow}r^{\downarrow*}$, where $r^{\uparrow}$ and $r^{\downarrow}$ are the reflection coefficients for scattered electrons, initially flowing from the normal metal towards the magnetic insulator. At a metal-insulator interface, the reflection probabilities are equal to one, but the reflection coefficients for spin-up and spin-down electrons have different phases, $r^\uparrow = exp(i\varphi_\uparrow)$ and $r^\downarrow = exp(i\varphi_\downarrow)$ [36]. When an electron with momentum $k$ enters the magnetic insulator of thickness $t$ with decay lengths $2/q_\uparrow$ and $2/q_\downarrow$ for spin-up and spin-down electrons, respectively, the phase difference of the reflection coefficients is however $\varphi_\uparrow - \varphi_\downarrow = ArcTan\left(\frac{Tanh(q_\uparrow t)k}{q_\uparrow}\right) - ArcTan\left(\frac{Tanh(q_\downarrow t)k}{q_\downarrow}\right)$. Thinner magnetic layers influence this phase difference in two ways. The first is the direct impact through the reduced thickness $t$ where the different exchange potential is experienced. When the magnetic insulator becomes thinner than the decay length, the electrons penetrate the magnetic insulator and reach the substrate (vacuum in our simple model), significantly reducing the phase difference and thus the efficiency of the SOT. Second, the larger atomic surface to volume ratio decreases the exchange interaction and Curie Temperature ($T_C$) [37] as well as the exchange splitting energy $2\Delta E_{ex}$. In studies on other magnetic insulators such as EuO, $2\Delta E_{ex}$ was shown to be proportional to $M_s$ [38]. As $M_s$ increases with $t_{TmIG}$ when $t_{TmIG} < 5.4$ nm, the exchange splitting energy for TmIG increases, yielding the increased spin mixing conductance and hence SOT efficiency $\xi_{eff}$ with increasing of the TmIG thicknesses. If the sample is thicker ($t_{TmIG} > 5.4$ nm), the exchange interaction reaches a bulk saturation value, $\varphi_\uparrow - \varphi_\downarrow = k/q_\uparrow - k/q_\downarrow$, and the interfacial torque is diluted in thicker samples, leading to the decrease of the SOT efficiency $\xi_{eff}$ through its simple $1/t$ dependence, which is consistent with our model.

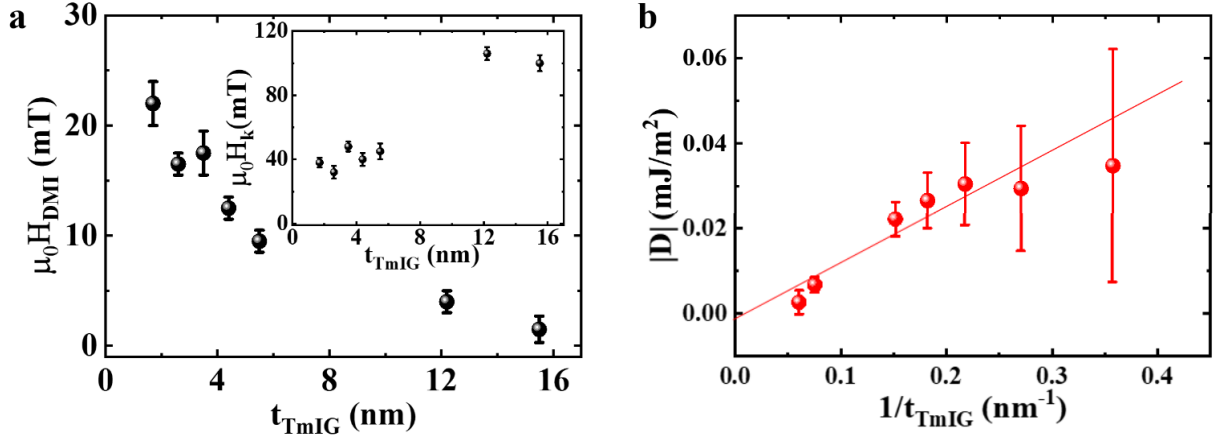

**Figure 3.** (a) TmIG thickness dependence of $\mu_0 H_{DMI}$ and $\mu_0 H_k$ (inset). (b)The effective DMI energy density $|D|$ as a function of $1/t_{TmIG}$, the linear relationship indicates an interfacial origin of the DMI.

The SOT determination discussed above are based on the current driven Néel domain wall motion, and the spin structure of these walls is stabilized by the DMI. As previously reported, the interfacial spin-orbit coupling in metals can induce both SOT and DMI at the FMM / HM interface [17], and in insulators the two effects can occur at different interfaces, in the case that the DMI is generated at the GGG / TmIG interface [21,26], We next study the DMI origin in GGG / TmIG / Pt, by means of the effective DMI energy density as a function of the TmIG and Pt thickness, to see if in this system the SOT and DMI are indeed originated at different interfaces. Figure 3a shows the TmIG thickness dependence of $\mu_0 H_{DMI}$ for GGG / TmIG ($t_{TmIG}$)/ Pt (7) samples. The $\mu_0 H_{DMI}$ decreases monotonically with increasing $t_{TmIG}$, which indicates the interfacial origin of the DMI. Together with the $t_{TmIG}$ dependence of $\mu_0 H_k$, we can calculate the DMI energy density as discussed previously [21]. The value of the $|D|$ scales linearly with the inverse of the TmIG thickness, as shown in Figure 3b, confirming the interfacial origin. Moreover, to conclude whether the GGG / TmIG or the TmIG / Pt interface is responsible for the interfacial DMI, we studied the $t_{Pt}$ thickness dependence of the DMI energy density, expecting that the Pt thickness would influence the DMI if this was generated at the TmIG / Pt interface [39].

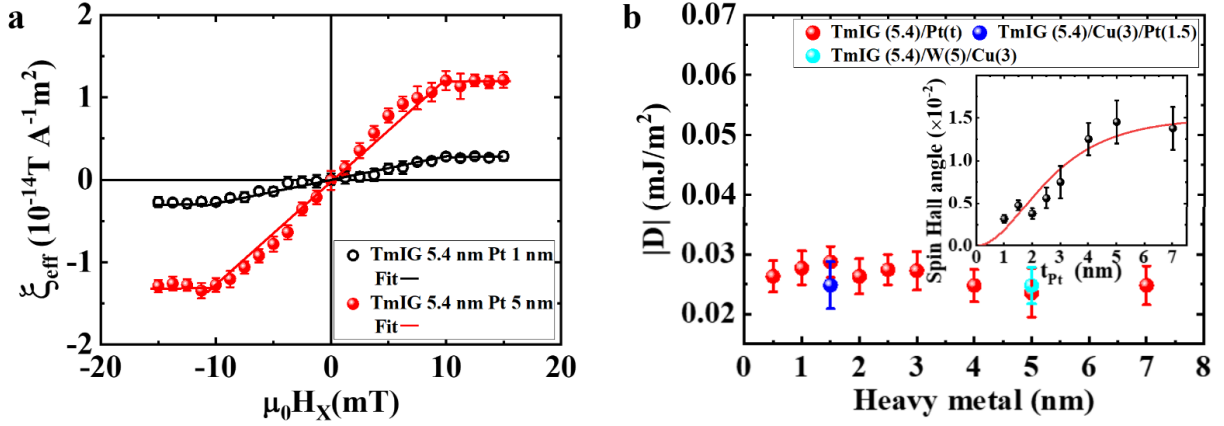

**Figure 4.** (a) Measured SOT efficiency $\xi_{eff}$ and $\mu_0 H_{DMI}$ for GGG / TmIG (5.4) / Pt (1) and Pt (5). $\xi_{eff}$ significantly depends on the Pt thickness, while $\mu_0 H_{DMI}$ is not affected. (b) The systematic comparison of the DMI energy density as a function of the HM thickness (Pt), HM material (W), and spacer layer (Cu). The magnitude of $|D|$ does not depend on the TmIG / HM interface (Note: The data for the 0.5 nm Pt sample is obtained from GGG / TmIG (5.4) / Pt (0.5) / Cu (3)) (units in nanometer).

Figure 4a shows the dependence of SOT efficiency $\xi_{eff}$ on the in-plane field $\mu_0 H_x$ for GGG / TmIG (5.4) / Pt (1) and GGG / TmIG (5.4) / Pt (5) samples, respectively. The torque efficiency $\xi_{eff}$ is smaller for the sample with smaller $t_{Pt}$ since the charge-to-spin conversion rate due to the SHE in Pt is reduced if the thickness is much smaller than the spin diffusion length. However the decrease of spin accumulation does not affect the value of $\mu_0 H_{DMI}$, we find $\mu_0 H_{DMI} = 10 \pm 1$ mT for both samples. Since the thicknesses of TmIG and the effective anisotropy field $\mu_0 H_k = 45 \pm 5$ mT are the same for both samples (details, see supplemental material S3 [40]), the same DMI energy density can be estimated. Furthermore, the Pt thickness dependence of $|D|$ in GGG / TmIG (5.4) / Pt ($t_{Pt}$) can be found in Figure 4b. If the iDMI originates from the ferromagnet / Pt interface, one would expect that |D| increases with increasing $t_{Pt}$, and reaches a saturation value at a thickness larger than the spin diffusion length [39]. Instead, we do not detect a significant variation of |D| with varying $t_{Pt}$, which indicates that the GGG / TmIG interface is the origin of the iDMI [41]. The effective spin Hall angle of GGG / TmIG (5.4) / Pt ($t_{Pt}$) can be found in the inset of Figure 4b. We can estimate the spin diffusion length $\lambda_{sf}$ in Pt, neglecting the spin relaxation caused by the Elliott-Yafet (EY) mechanism, by fitting with the function: $\theta(t_{Pt}) = \theta(\infty)(1 - sech(t_{Pt} / \lambda_{sf}))$ [42], which yields $\lambda_{sf}$ =1.8 ± 0.3 nm.

To further check if the GGG/ TmIG interface is the source of the iDMI, we included a 3 nm Cu spacer layer between Pt and TmIG to make a GGG / TmIG (5.4) / Cu (3) / Pt (1.5) structure. Cu is a typical light metal with a long spin diffusion length and small spin-orbit coupling. As shown in Figure 4b, inserting a Cu spacer does not strongly affect the DMI energy density. Hence, direct contact with HM and proximity effect are not the main reasons for the iDMI [26]. Replacing the Pt with W also induces little changes in the iDMI shown in Figure 4b, which indicates that the HM is not responsible for the iDMI [43]. This again shows that the iDMI is majorly determined by the GGG / TmIG interface in this stack, which is not the interface where the SOTs originate from.

In summary, we have performed a systematic thickness dependence study to reveal the origin of the SOT and DMI in GGG / TmIG / Pt heterostructures. We quantify the SOT efficiency as a function of the TmIG thickness, revealing a non-monotonic dependence. We can explain our results by considering that the SOT is an interfacial effect in this system, but the spin mixing conductance of the TmIG / Pt interface is affected by the thickness-dependent exchange splitting energy and the resulting spin current penetration depths in TmIG. The reduction of the spin-torque efficiency at low TmIG thickness is correlated with the reduction of the saturation magnetization in the thin films and both originate from a thickness-dependent exchange interaction. Comparing SOTs and DMI, we find from the TmIG and Pt thickness-dependence of the DMI energy that the GGG / TmIG interface is responsible for the iDMI. Our results show that the thickness-dependent properties of the ferromagnetic insulators have a strong influence on the SOT and DMI that are crucial to manipulate the domain walls and magnetization electrically. Moreover, these results pave the way towards the control of the chiral magnetic textures in insulating systems, where the SOT and DMI efficiencies can be independently controlled by tuning the properties of the different interfaces.

**Corresponding Author**

Corresponding Authors

*E-mail: klaeui@uni-mainz.de

**Acknowledgments:**

We acknowledge support from the Graduate School of Excellence Materials Science in Mainz (MAINZ) DFG 266, the MaHoJeRo (DAAD Spintronics network, Project No.57334897), Deutsche Forschungsgemeinschaft (DFG, German Research Foundation) – Spin+X TRR 173 – 268565370 (project B02), DFG project 358671374, the Research Council of Norway (QuSpin Center 262633), ERATO "Spin Quantum Rectification Project" (Grant No. JPMJER1402), National Key Research and Development Program of China (Grant No. 2016YFB0700901, 2017YFA0206303, 2017YFA0403701), and National Natural Science Foundation of China (Grant No. 51731001, 11675006, 11805006, 11975035). L.B acknowledges the European Union's Horizon 2020 research and innovation program under the Marie Skłodowska-Curie grant agreements ARTES number 793159. A.R acknowledge the Max Planck Graduate Center with the Johannes Gutenberg-Universität Mainz (MPGC).